# The Possible Effects of Personal Income Tax and Value Added Tax on Consumer Behaviors

Dr. Ahmet AK

Professor Faculty of Economics and Administrative Sciences Department of Public Finance, Ankara Haci Bayram Veli University Turkey

Email: ahmet.ak@hbv.edu.tr (Author of Correspondance)

Oner Gumus

Research Assistant, Kutahya Dumlupinar University Turkey

Email: oner.gumus@dpu.edu.tr

**Abstract**

In economics literature, it is accepted that all people are rational and they try to maximize their utilities as possible as they can. In addition, economic theories are formed with the assumptions not suitable to real life. For instance, indifference curves are drawn with the assumptions that there are two goods, people are rational, more is preferred to less and so on. Hence, the consumer behaviors are guessed according to this analysis. Nevertheless, these are invalid in real life. And this inconsistencey are examined by behavioral economics and neuroeconomics. Behavioral economics claims that people can behave what they are not expected since people can be irrational, their willpower is limited and altruistic behaviors can be seen and they can give more value to what they own. As a result of these, consumer behaviors become more different than that of economic theory. In addition to behavioral economics, neuroeconomics also examines consumer behaviors more differently than mainstream economic theory. It emphasizes the people using prefrontial cortex of the brain are more rational than the people using hippocampus of the brain. Therefore, people can make illogical choices compared to economic theory. In these cases, levying taxes such as personal income tax or value added tax can be ineffective or effective. In other words, the effect becomes ambigious. Hence, the hypothesis that if government desires to levy personal income tax or value added tax, it makes a detailed research in terms of productivity of taxes forms the fundamental of this study.

## 1. Introduction

Every choice is not necessarily suitable to rationality. In addition, there are lots of situations seen around the world which are irrational. For example, even though some people are fat, they do not give up eating fatty foods or people can choice altruism instead of saving or they can give more value to what they own. However, mainstream economy do not accept this; rather, it makes assumptions which are not realistic in real life. Most of the analyses are based on these unrealistic assumptions. That's why, in first part we emphasize the consumer choice theory to show how mainstream economy sees the consumer behaviors.

In the second part, two popular economic approaches -behavioral economics and neuroeconomics are told. Because the people behave irrationally in real life, these two approaches try to understand why they behave like this.

In the third part, the theoretical foundations of personal income tax and value added tax is are presented. Personal income tax decreases the disposable income and value added tax increase the price of the goods and services. Thus, it cannot be said that they are not effective on consumer choices.

In the last part, general assesments are made in terms of behavioral economics and neuroeconomics and the effects creating by the two taxes. And the study is ended with the conclusion part which states that the government ought to be careful while levying personal income tax and value added tax.

## 2. Consumer Choice in Economics

The people have to live as a whole and all activities of people aims to meet the needs [1]. Need is a humane feeling and people are under such a pressure during their lives and if this need is met, they feel and gain pleasure; If not, they rue and feel pain [1]. For instance, if a thirsty person do not find water, they feel pain; if he/she finds water, they gain pleasure due to meeting his/her feeling of thirst [1].

The features of needs are as follows [2]:

- Needs are continuous,
- Needs show a tendency to increase,
- Primitive needs lost its violence if they are met.

To meet needs, one can make a consumption. Consumption can be divided into two categories [3]:

- **Productive Consumption:** The consumption which creates production is called productive consumption. Every production requires a consumption. For instance, bread production completely requires flour, salt and wood to consume and partly requires building and tools to consume. The reason why we say this kind of consumption is productive is that the benefit produced is more than the benefit consumed.

- **Unproductive Consumption:** If the consumption is just related to meet our needs, it is called unproductive consumption. Consumption of food, wearing clothes, dwelling in a house, furnishing this house with furnitures are examples of unproductive consumption. Because there is no production activity in this kind of consumption, we say this as unproductive consumption.

Despite of these kind of consumptions, it is not suitable to divide the population as producer or consumer since every producer is consumer at the same time and every consumer is producer at the same time [4]. For instance an officer produces public service while he/she is consuming something for his/her livelihood and while a farmer produces wheat, this farmer consumes every kind of goods and services [4]. When it is told consumer population, it means the people who are not active; so, this group generrfally consists of children and the old meaning that children are producers of future and old people are producers of past [4]. It does not matter whether there is a production or consumption, If it is produced or consumed, the main point is good. Hence, it ought to be examined carefully.

Goods are divided as free goods and economic goods; economic goods can be divided as the goods produced by nature and goods produced by people; and the goods produced by people can be divided as production goods and consumption goods [5]. Nevertheless, people have to make choice when buying something since there is scarcity. To examine this, the indifference curves are used.

Indifference curves demonstrate the combinations of consumption bundles giving the consumer the equal utility [6]. Indifference curves have the following properties [7]:

- Every point on the indifference curve shows the combinations of consumption providing equal total utility,
- The point which is the farthest from the origin gives the consumer the highest utility,
- Indifference curves cannot cross each other,
- Indifference curves have negative slopes,
- Indifference curves are convex to the origin.

Based on these properties, five principles can be said concerning indifference curves [8]:

- **Limited Income Requires Preference:** Limited income forces consumers to make preference concerning which goods they are going to buy anda re not going to buy.
- **Consumers Have An Aim When Making Decision:** Because it is accepted that consumers are rational, they choose the highest benefit good by making cost-benefit analysis.
- **There Can Be Substitute Goods:** This is related to that consumers can satisfy many available different alternatives and thereby achieving utility.

- **Even Though There Is No Perfect Information Knowledge and Past Experiences Are Going to Help Consumers When They Have To Make Decisions:** The good decision is based on the value of the good the consumers give. By this way, they can make a good choice.
- **The Law of Diminishing Marginal Utility is Valid:** This means that the marginal utility obtained by means of consumption of successive units of a product is going to decrease while the rate of consumption increases.

When consumers make chooice between two goods, they face budget constraint.

### Graph 1. Budget Constarint

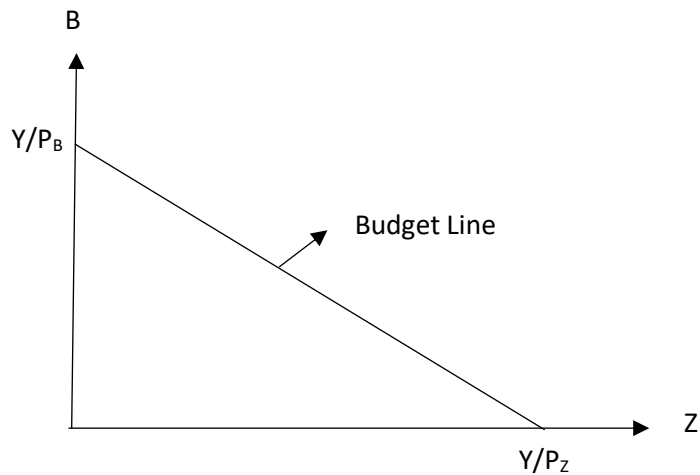

**Source:** The graph is drawn by the authors by inspiring [9].

Budget constraint can be defined by a formula $P_B.B+P_Z.Z=Y$, where $P_B$ is the price of good B, B is the amount of good B, $P_Z$ is the price of good Z, Z is the amount of good Z and Y represents income [9].

When looked Graph-1, it can be seen that two events can change the budget line [10]:

- A change in income results in parallel shifts in the budget line,
- A reduction in price of good B causes budget linet o move outward along the axis representing the quantity of good B while leaving the location of good Z unchanged.

It can be examined the income and substitution effects based on the changes in the price of one good. They can be defined as follows [11]:

- **Income Effect:** The decrease in the price of one good causes more of two goods to purchase. This is called income effect.
- **Substitution Effect:** The decreasein the price of one good causes more of this good to purchase. This is called substitution effect.

*The Possible Effects of Personal Income Tax and Value Added Tax on Consumer Behaviors*

All of these analyses associated with consumer choice are based on some assumptions. In real life, however, these assumptions can be ineffective. The reasons are:

- In the analyses, there are two goods. In real life, nevertheless, consumers can face many of goods and services. In that case, the analysis cannot be valid.
- In economics literature, there is a term which is called scarcity. But, sometimes people can exceed this phenomenon by using illegal ways. Suppose that a leader of illegal group can create economic resources by violating the law. In that case, all the needs of this illegal person are met. By this way, this analyses cannot be valid.
- In this analyses, a good is accepted as normal good. However, a good can also be giffen good or inferior good. Hence, the case changes.
- An individual can be addictive of something thereby causing indifference curve to be concave to the origin. So, the analysis can change.

These events can be accounted for by using behavioral economics and neuroeconomics.

**2. Behavioral Economics and Neuroeconomics**

Behavioral economics is the struggle to rise the explanatory and predictive power of economic theory by enabling it with more psychologically logical foundations and historically, it was a direct result of cognitive revolution [12].

There are six principles in behavioral economics [13]:

- Although people try to prefer the best logical option, they sometimes become unsuccessful,
- People are partly engaged in concerning how their conditions compare to referance points,
- People experience self-control problems,
- People are interested in actions, intentions and payoffs of other goods or services as well as the payoff of the good or service they desire to purchase,
- A lot of psychological factor can be important for market,
- Theoretically restraining people's choice partially save them from behavior bias; nevertheless, paternalistic behavior of governments have bad experiences and often unpopular.

In behavioral economics there are three obstacles hampering rational choice [14]:

- **Bounded Rationality:** This is related to limitation of computing power of the human brain.
- **Bounded Willpower:** This is associated with the choice we firstly make, but we regret later.
- **Bounded Self-Interest:** This is related to a suppression that forces us to help people instead of maximizing our interest.

In addition, there is endowwment effect a behavior which is more general and may affect the choices [14]. It means there is a tendency for consumers to value something more highly since they own it [14].

Behavioral economists found something concerning behavioral economics [15]:

- People confident theirselves mushily,
- People give too much value to a small number of exact observations,
- People are unwilling to do change their minds.

By using the information, one can understand that there is always a vivid possibility for people to make rational choice. However, there is another branch of economics examining consumer choices which is called neuroeconomics.

Neuroeconomics uses the sciences of neuroscience, psychology, economics and computer science to search three fundamental questions [16]:

- To make different types of decisions what are the variables computed by the brain and how are they associated with the outcomes as a result of the behavior an individual shows?
- How does the underlying neurobiology apply and force these computations?
- What are the results of this knowledge for understanding behavior and well being in miscellaneous contexts- economic, policy, clnical, legal, business and others?

Economics and neuroscience have to learn a lot from one another and the gallant of this interpretation is neuroeconomics meaning the application of neuroscientific methods to analyze and understand economically relevant behavior [17]. The link between neural mechanisms and choices is supposably to follow the three kinds of support and inspiration for areas of economic theory [18]:

- Neural evidence of the utility maximization in simple choice,
- Neural evidence severalizing different behavioral and rational processes,
- Neural evidence of other psychological effects.

Neuroeconomics examines rational and irrational choices as follows [14]:

- People store memories, analyze data and expect the results of their actions in prefrontal cortex of the brain where rational utility mazimization occurs,
- People store memories of anxiety and fear in the hippocampus of the brain where the decision is made as non-rational driven by fear or anxiety.

## 3. Personal Income Tax and Value Added Tax

The theoretical features of personal income tax are as follows [19]:

- It is a direct tax,
- Because it is levied on the income of individuals the incidence of this tax is very difficult,

- It is a subjective tax since it can be harmonized based on taxpayers' personal, familial, and social structure,
- Because expenditures can be extracted from gross income, the net income is taxed,
- Self consumption is stayed out of taxation,
- Because it is a direct tax, it has a negative effect on savings and thereby decreasing the funds for investments,
- Personal income tax creates a decrease in purchasing power and thereby decreasing demand. As a result, the economy can experience a recession.

Three systems are applied in income taxes [20]:

- **Schedular Tax Incomes:** The incomes are determined as seperate schedules and every schedule is processed seperately in terms of tax,
- **Total Tax Incomes:** They have subjective characters. All the incomes taxpayers earn from various sources are considered. The amounts obtained by considering taxpayer's personal and familial situations are extracted from this total income. The amount left after the tools reaching ability to pay are applied is taxed.
- **Mixed Tax Incomes:** Income factors are taxed as schedular in first step, after that these amounts are added and this total amount are taxed again.

Value added is the difference between buying cost and selling price of the goods and services businesses sell in terms of business; namely, value added brings the material which business purchases as intermediate good incremental value and this value is total payment (wage, interest, rent, depreciation, profit) businesses make for the factors of production for the good produced in the level of production [21]. Therefore, value added means total amount of economic values [22].

In value added tax, there is a delivery of goods and services as a result of the activities like commercial, industrial, agricultural, import and so on [23]:

- **Delivery of Goods:** It is transfer of the right of disposition to another one by the agents who moves on behalf of owner of goods.
- **Delivery of Services:** It is related to the fulfillmentof service against a remuneration.

## 4. General Assessments of Consumer Choice in Behavıral Economics and Neuroeconomics: The Perspectıve of Taxatıon

When inflation rises, governments generally levy taxes to put the fire out. In that case, the price of goods and services increases more than that of past. According to law of demand, the demand for goods and services decreases. This is functional for people who are rational and use prefrontial cortex of the brain.

Nevertheless, sometimes the kind of good can be important. For instance, if the good is a giffen good, the demand for giffen good increases.

If the consumer is a person who uses hippocampus of the brain, this means that the consumer is irrational. Such kind of consumers make illogical choices. For instance, even though this consumer knows that hamburger is damaged to the health, he/she can buy it due to his/her addiction. In other words, the elasticity of demand for hamburger is inelastic fort his customer. In that case, despite of an increase in personal income tax on customer's income or value added tax on hamburger, this consumer behaves irrationally.

Suppose that there is a heavy smoker. Although he/she knows that smoking is damaged in terms of money and health, this consumer can buy it. In this case, because of inelastic demandi the consumer is not interested in whether government levies personal income tax or value added tax. Therefore one can say that this consumer behaves irrationally.

Think a person going on a diet and assume that government levies a personal income tax. This person can stop the diet and buy something fatty even tough he/she regrets later. By this way, one can say that this person uses hippocampus of the brain instead of prefrontal cortex of the brain.

Some altruistic behaviors can deviate from being rational. Suppose government levies value added tax on goods and services menaing that consumers can purchase less than before. Despite of this fact, these consumers can go on helping the poor.

## 5. Conclusıon

Economic theory always says that resources are scarce and people always run for their inetersts to maximize their utilities. In addition, it emphasizes the people are rationalwhen reaching the maximum utility. For example, if government levies personal income tax, the disposable income people decreases. A rational people using prefrontial cortex of the brain do not consume; rather, he/she will be prone to save. However, sometimes people are not be able to behave rationally. In addition, their willpower can be low or altrustic behaviors become more important than being rational. In these cases, people are inclined to use hippocampus of the brain rather than prefrontal cortex of the brain. In other words, they can diverge from being rational. So, such kind of events can change the expected effects of taxation (personal income tax and value added tax) and levying the taxes become weaker or stronger-namely the effect of tax can be ambigious. That's why, a government desiring levying personal income tax and value added tax should be considered carefully because in real life, society can consist of more people using hippocampus than the people using prefrontal cortex or vice versa.

*The Possible Effects of Personal Income Tax and Value Added Tax on Consumer Behaviors*

## Author Details

**Dr Ahmet AK** is a professor of Faculty of Economics and Administrative Sciences Department of Finance, Ankara Haci Bayram Veli University. He received his Ph.D. Anadolu University (Public Finance/Tax Law) Eskişehir, 2004 (Ph.D. Dissertation, Tax Law Drafting Legislation Process and Practice in Turkey). He is very interested in research mostly on Social Science and Humanities, Public finance, Taxation &Tax Law, Religious

Studies (especially Islamic finance). Ahmet AK is working with various reputed journals as editor, co-editor and reviewer.

**Author Details**

**Öner Gümüş** was born in Antakya/TURKEY in 1984. He graduated from Anadolu University the department of English Economics (2010), the department of Public Finance (2015) and the department of Jurisprudence (2018) and he has master degree on Economics and Public Finance. He is going on his studies in Dumlupınar Unicersity Social Sciences Institute Public Finance PhD Program as research assistant. He has lots of scientific studies published in national and international journals, and publishing companies.